\documentclass{article}

\usepackage[margin=1.1in]{geometry}
\usepackage{amsmath}
\usepackage{amsfonts}
\usepackage{amssymb}
\usepackage{graphicx}
\usepackage{caption}
\usepackage{subcaption}
\usepackage{mathtools}
\usepackage{natbib}
\usepackage{hyperref}
\usepackage[capitalise]{cleveref}
\usepackage{fancyhdr}

\graphicspath{{Figs/}}

\newcommand{\mathsfbi}[1]{\boldsymbol{\mathsf{#1}}}
\newcommand{\pder}[2]{\frac{\partial #1}{\partial #2}}

\newcommand{\Dderfull}[1]{\pder{#1}{t}+u\pder{#1}{x}+w\pder{#1}{z}}
\newcommand{\pderline}[2] {{\partial #1}/{\partial #2}}
\newcommand{\dint}{\,\textrm{d}}

\renewcommand{\v}[1]{{\bf #1}}
\newcommand{\mat}[1]{\mathsfbi{#1}}

\newcommand{\E}{\mathrm{E}}
\newcommand{\Ro}{\mathrm{Ro}}

\DeclareMathOperator{\erf}{erf}

\title{\bf Can vertical mixing arrest ocean frontogenesis?}
\author{Matthew N. Crowe\footnote{Matthew.Crowe2@ncl.ac.uk}$\hspace{5pt}^{1,2}$ \\
\small{$\,^{1}\,$School of Mathematics, Statistics and Physics, Newcastle University, Newcastle upon Tyne, NE1 7RU, UK} \\ \small{$\,^{2}\,$Department of Mathematics, University College London, London, WC1E 6BT, UK}}
\date{}
\date{}

\begin{document}

\maketitle

\begin{abstract}
Frontogenesis is the process by which ocean fronts--regions of strong horizontal density gradients in the upper ocean--strengthen.  It is a highly non-linear phenomenon arising from many competing effects, such as background strain, instabilities, and surface forcing. Vertical mixing plays a complicated role in governing both frontogenesis and the long-term dynamics of ocean fronts. For weak fronts, it has been shown to drive shear dispersion which further weakens the front over long timescales. However, vertical mixing also mixes the vertically-sheared thermal wind velocity, driving the front out of balance and causing a secondary circulation to develop that may strengthen the front. Here, it is shown that vertical mixing can directly act to oppose the frontogenetic secondary circulation, provided the frontal width is larger than a critical value. A non-linear iteration method is described for finding balanced states in which vertical mixing arrests frontogenesis and the parameter dependence of this regime is discussed. Further, existing asymptotic solutions for the buoyancy and velocity fields are extended and found to closely match these balanced states, even outside their region of formal validity.
\end{abstract}

%\keywords{1,2,3}

\section{Introduction}

Ocean fronts are regions of elevated horizontal density gradient in the upper ocean. They are dynamically active regions, associated with strong bands of vertical velocity and hence  important for the vertical transport of tracers and gas exchange between the ocean and the atmosphere \citep{GARRETTLODER,FERRARI}. Ocean fronts coexist with strong turbulent mixing due to various surface processes; such as surface heat flux and wind stress, ocean-ice interactions, and submesoscale instabilities. The interplay between ocean fronts and turbulent mixing has been the topic of many recent works and will be the primary focus of this paper.

Fronts exist in a state close to thermal wind balance; the balance between horizontal pressure gradients arising from the front and the Coriolis force associated with along front, vertically-sheared jets. Various processes---such as background strain and inertial transients \citep{HOSKINSBRETHERTON,BLUMEN,SHAKESPEARETAYLOR}---can drive a front out of thermal wind balance, leading to re-stratification via a secondary circulation and the sharpening of surface density gradients. This sharpening---referred to as frontogenesis---can lead to the formation of a surface discontinuity in the density field in the inviscid limit. Such infinitely sharp fronts are not observed physically, suggesting the presence of a mechanism opposing this sharpening. It is generally believed that frontogenesis is opposed by turbulent mixing processes in the surface mixed layer \citep{DauhajreEtAl2025}. In many numerical models, parametrised dissipative mixing is included \citep{LARGEETAL}, preventing fronts collapsing to grid-scale singularities. However, the physical validity of such parametrisations is often poorly understood.

Turbulent mixing plays a complicated role in the dynamics of ocean fronts. Strong turbulence mixes the vertically sheared thermal wind jet, driving the front out of balance and creating a secondary circulation which acts to re-stratify the front \citep{THOMPSON} and may lead to frontogenesis \citep{MCWILLIAMS,BodnerEtAl2019}. Simultaneously, the combination of a vertical stratification and a vertically-sheared cross-front flow leads to shear dispersion, where vertical diffusivity is projected horizontally, leading to frontolytic spreading \citep{YOUNG,CROWETAYLOR,CROWETAYLOR3}. This process is further complicated by the coupling between the frontal structure and the strength of the turbulent mixing. As sharp gradients form in the density field, various instabilities can occur \citep{DauhajreEtAl2025,AtkinsonEtAl2025}, resulting in an enhanced turbulent mixing which may feed back onto the frontal structure.

Typically, the vertical length-scales of ocean fronts are much smaller than their cross-front scale, so vertical mixing processes play a more important role than horizontal ones. This assumption leads to turbulent thermal wind (TTW) balance: a quasi-steady three-way balance between cross-front pressure gradients, the Coriolis force, and the vertical mixing of momentum which has been found to approximately hold in both numerical models and observations \citep{CRONINKESSLER,TAYLORFERRARI,WENEGRATMCPHADEN}. Formally, this balance arises in the small Rossby number limit for a front in the presence of strong vertical mixing, requiring a weak front and hence a large cross-front length-scale. \citet{CROWETAYLOR} derived analytical TTW solutions for small Rossby number and argued that many aspects of TTW balance are valid even when the Rossby number is not small \citep{CROWETAYLOR3}. The validity of this balance in cases with finite Rossby number will be further addressed here.

The primary goal of this study is to isolate the effects of vertical mixing on an idealised ocean front in the case of finite Rossby number. These is done by seeking non-linear steady state solutions \citep{CessiIerley}, thereby ignoring the effects of instabilities and inertial transients. The frontolytic effects of shear dispersion \citep{CROWETAYLOR,CROWETAYLOR3,YOUNG} are similarly neglected by removing the depth-averaged components which drive shear-dispersive spreading. This allows us to determine the extent to which vertical mixing alone can prevent the formation of a surface discontinuity in the density field and arrest frontogenesis. We begin in \cref{sec:setup} by describing the problem setup and underlying assumptions. We then discuss the non-linear iteration method employed to find steady-state solutions in \cref{sec:method}. In \cref{sec:results} we present our results and describe an extension to the TTW theory of \citet{CROWETAYLOR}. Finally, we give our conclusions in \cref{sec:conc} and discuss some avenues for future work.

\section{Problem Setup}
\label{sec:setup}

Consider a three-dimensional region of fluid of depth $H$, bounded above and below by rigid boundaries. We assume that the fluid is rotating about the vertical ($z$) axis with rotation rate $f/2$, for Coriolis parameter $f$. Density variations are represented by a buoyancy field, $b$, and we denote cross-front position by $x$ and along-front position by $y$. The velocity and pressure fields are denoted by $\textbf{u} = (u,v,w)$ and $p$ respectively. Along-front variations on buoyancy are assumed to be small, hence we neglect all $y$ derivatives, while allowing for a $y$-independent along-front velocity $v$.

The cross-front length-scale is taken to be the frontal width, $L$, which is assumed to be much larger than $H$ so the aspect ratio is small, $\delta = H/L \ll 1$. Vertical mixing is represented by a viscosity $\nu$ and diffusivity $\kappa$ and assumed to dominate over horizontal mixing. We allow $\nu$ and $\kappa$ to depend on the depth $z$, $(\nu, \kappa) = \big(\nu(z), \kappa(z)\big)$, for consistency with common mixed layer parametrisations \citep{LARGEETAL} and previous studies \citep{CROWETAYLOR}, but neglect horizontal variations for simplicity.

The nondimensional governing equations are
\begin{subequations}
\label{eq:TTW}
\begin{alignat}{3}
\label{eq:TTW_a}
\Ro\left[\Dderfull{}\right]u - v = & -\pder{p}{x}+\pder{}{z}\left[\E(z)\pder{u}{z}\right],\\
\label{eq:TTW_b}
\Ro\left[\Dderfull{}\right]v + u = & \hspace{39pt} \pder{}{z}\left[\E(z)\pder{v}{z}\right],\\
\label{eq:TTW_c}
\delta^2\Ro\left[\Dderfull{}\right]w - b = & - \pder{p}{z} +\delta^2 \pder{}{z}\left[\E(z)\pder{w}{z}\right],\\
\label{eq:TTW_d}
\Ro\left[\Dderfull{}\right]b \hspace{20pt} = & \hspace{38pt}\pder{}{z}\left[\frac{\E(z)}{\Pr(z)}\pder{b}{z}\right],\\
\label{eq:TTW_e}
\pder{u}{x}+\pder{w}{z} \hspace{60pt} = & \quad 0,
\end{alignat}
\end{subequations}
where the Rossby number, $\Ro = \delta \Delta b/(f^2 L)$, describes the strength of non-linear momentum advection relative to the Coriolis force, and $\Delta b$ is (half) the buoyancy jump across the front. Nondimensional vertical mixing is represented by an Ekman number, $\E(z) = \nu(z)/(fH^2)$, and Prandtl number, $\Pr(z) = \nu(z)/\kappa(z)$ which may depend on depth, $z$. The $O(\delta^2)$ non-hydrostatic terms in \cref{eq:TTW_c} are retained for generality, but found to have no significant on the behaviour of the system for the parameter ranges considered here.

Boundary conditions of no stress, no buoyancy flux and no vertical velocity are imposed at the top and bottom surface so
\begin{equation}
\label{eq:BCs}
\pder{}{z}(u,v,b) = \textbf{0}\quad \textrm{and}\quad w = 0 \quad \textrm{on} \quad z = \pm \frac{1}{2}.
\end{equation}
Finally, the front is represented by a background buoyancy field
\begin{equation}
\label{eq:b0_def}
b_0(x) = \erf\left[\frac{\!\sqrt{\pi}}{2} x \right],
\end{equation}
where $\erf(x)$ denotes the error function. The factor of $\sqrt{\pi}/2$ is chosen so that the horizontal frontal gradient, $\pderline{b_0}{x}$, has a maximum value of $1$ within the front. In this framework, the Rossby number, $Ro$, may be thought of as the inverse frontal width, with large Rossby numbers corresponding to small cross-front scales.

\section{Methodology}
\label{sec:method}

We aim to determine if vertical mixing alone is sufficient to prevent frontogenesis in the case of finite Rossby numbers. This is done by seeking non-linear steady state solutions to \cref{eq:TTW}. If such steady state solutions exist, we can conclude that vertical mixing may prevent the formation of surface singularities, and enable equilibria to exist where the frontogenetic secondary circulation is balanced by frontolytic vertical mixing processes.

An advantage of this approach is that is enables us to ignore various other frontogenetic and frontolytic effects and isolate the vertical mixing processes of interest. In particular, neglecting the time derivatives allows us to ignore inertial transients---which may drive unrealistic frontogenesis \citep{AtkinsonEtAl2025}---and the complicated effects of frontal instabilities. Since the frontolytic, shear-dispersive effects of vertical mixing through the depth-averaged buoyancy field are well understood \citep{YOUNG,CROWETAYLOR,CROWETAYLOR3}, we subtract the depth-average components from \cref{eq:TTW} and consider only the deviations from the depth-average. Unlike the depth-averaged components which continually evolve through slow, shear-dispersive spreading of the front, steady state solutions are expected to exist for the deviations due to the higher number of degrees of freedom. If vertical mixing is sufficient to prevent frontogenesis for a given state without shear dispersive effects, it is also expected to prevent frontogenesis once these (frontolytic) effects are included. Further, the effects of shear dispersion are known to be much slower that the formation of a secondary circulation \citep{CROWETAYLOR, CROWETAYLOR3}, so are unlikely to play a significant role in preventing frontogenesis for strong fronts.

%The above assumptions may be justified by noting that, even for finite $\Ro$, the depth-averaged buoyancy and vorticity fields evolve slowly compared to the timescale of secondary circulation evolution, so a system initialised with a specific $b_0$ and no depth-averaged velocity will remain close to this state while the depth-dependent fields evolve. 

%First, it should be noted that true steady state solutions to \cref{eq:TTW} are unlikely to exist as it was shown in \citet{CROWETAYLOR,CROWETAYLOR3} that over long timescales, the effect of vertical mixing is to spread a front via shear dispersion.
%As this spreading drives a weakening of cross-front buoyancy gradients, it is purely frontolytic.

%Here, the aim is to study how frontogenetic effects arising from finite Rossby numbers can be directly opposed by vertical mixing.
%Therefore, frontolytic spreading can be neglected as it will not enhance the frontogenetic circulation.

We proceed by neglecting all time derivatives from \cref{eq:TTW}, subtracting the depth-averaged form of \cref{eq:TTW} from the system, and considering only deviations from this depth-average. We define the deviation from the depth-average as
\begin{equation}
\mathcal{A}[f] = F - \int_{-1/2}^{1/2} F \dint z,
\end{equation}
and apply the operator $\mathcal{A}$ to \cref{eq:TTW}. Any terms corresponding to a depth-average of $b$ are set to the depth-independent background buoyancy, $b_0$, and any terms corresponding to a depth-average of the horizontal velocity fields are set to zero. The resulting system may be written as
\begin{equation}
\label{eq:BGS_mat}
\mat{N}\v{q} = \v{f},
\end{equation}
where
\begin{equation}
\mat{N} = \begin{pmatrix}
\mathcal{N}_u & -1 & 0 & 0 & \pder{}{x} \\
1 & \mathcal{N}_u & 0 & 0 & 0 \\
0 & 0 & \delta^2\mathcal{N}_u & 1 & \pder{}{z} \\
\Ro\pder{b_0}{x} & 0 & 0 & \mathcal{N}_b & 0 \\
\pder{}{x} & 0 & \pder{}{z} & 0 & 0
\end{pmatrix}, \quad \v{q} = \begin{pmatrix}
u \\ v \\ w \\ b \\ p
\end{pmatrix}, \quad \v{f} = \begin{pmatrix}
-z\pder{b_0}{x} \\ 0 \\ 0 \\ 0 \\ 0
\end{pmatrix},
\end{equation}
for the (non-linear) advection diffusion operators
\begin{equation} %\mathcal{A}\left[ \right)
\mathcal{N}_u \phi = \Ro\left(\mathcal{A}\pder{}{x}\left[u \phi \right]+\pder{}{z}[w\phi]\right)-\pder{}{z}\left[ \E \pder{\phi}{z}\right],
\end{equation}
and
\begin{equation}
\mathcal{N}_b\phi = \Ro\left(\mathcal{A}\pder{}{x}\left[u \phi \right]+\pder{}{z}[w\phi]\right)-\pder{}{z}\left[ \frac{\E}{\Pr} \pder{\phi}{z}\right].
\end{equation}
Here the fields, $(u,v,b,p)$, denote the depth-dependent component only and depth-average to zero. The full buoyancy, including the depth-independent background profile, is given by $b_0+b$. The full pressure is given similarly by $p+z\,b_0$, and the full horizontal velocities are depth-independent by assumption. The vertical velocity, $w$, has non-zero depth average and is uniquely determined by mass conservation.

\cref{eq:BGS_mat} may be discretised using spectral collocation \citep{Spectral_Methods}. We use (periodic) Fourier points in the $x$ direction and Chebyshev points of the second kind in the $z$ direction. The $x$ and $z$ derivatives may be expressed as matrices acting on vectors of function values and the full system converted into a matrix problem where the solution is represented by a state vector. For $N_x$ gridpoints in the $x$ direction and $N_z$ gridpoints in the $z$ direction, the discretised form of $\mat{N}$ is a $(5N_x N_z) \times (5N_x N_z)$ matrix denoted by $\tilde{\mat{N}}$ and the discretised forcing, $\v{f}$, and state vector, $\v{q}$, are vectors of length $5N_x N_z$ denoted as $\tilde{\v{f}}$ and $\tilde{\v{q}}$ respectively. The boundary conditions from \cref{eq:BCs} are incorporated into the discretised system
\begin{equation}
\label{eq:Disc_BGS}
\tilde{\mat{N}}\tilde{\v{q}} = \tilde{\v{f}},
\end{equation}
by replacing the rows corresponding to $z = \pm 1/2$ with the boundary conditions. \cref{eq:Disc_BGS} cannot simply be inverted to find $\tilde{\v{q}}$ since $\tilde{\mat{N}}$ depends of $u$ and $w$ through the advection operators. Instead, we define an initial solution $\tilde{\v{q}}_0$ using the small $\Ro$ asymptotic solution from \citet{CROWETAYLOR} and perform the iteration
\begin{equation}
\tilde{\v{q}}_{n+1} = \tilde{\mat{N}}_n^{-1} \tilde{\v{f}},
\end{equation}
where $\tilde{\mat{N}}_n$ is the matrix $\tilde{\mat{N}}$ evaluated using $u$ and $w$ from $\tilde{\v{q}}_{n}$. The squared difference in a given field between consecutive iterations is defined as
\begin{equation}
\Delta_n[f] = \int_D (f_n-f_{n-1})^2 \dint D,
\end{equation}
where $D$ denotes the 2D numerical domain. The scheme is iterated until either the total iteration difference is below a specified error tolerance
\begin{equation*}
\epsilon^2 > \Delta_n[u] + \Delta_n[v] + \Delta_n[b],
\end{equation*}
or the iteration fails to converge. As well as enabling us to neglect unwanted processes, this iteration procedure is fast---allowing for a large parameter sweep---and does not require horizontal viscosity/diffusivity for stability, allowing us to isolate just the effects of vertical mixing.

Due to the exponential accuracy of spectral collocation for smooth functions, only a small number of grid-points are required. Here we present results for $(N_x, N_z) = (64, 61)$, however additional iterations have been performed using different grid sizes to verify that our results are grid-independent. An error tolerance of $\epsilon^2 = 10^{-10}$ is used throughout and found to be sufficient to identify converged iterations. Here, we present and discuss results for constant Ekman and Prandtl numbers. However, results with depth-dependent $\E$ and $\Pr$ have been obtained and found to be quantitively similar to the constant case, with the magnitude of $\E$ and $\Pr$ being the most important factor in determining the behaviour of the system.

A converged solution represents a non-linear steady state solution to the finite Rossby number system. However, failure to converge does not necessarily imply that a state state does not exist for the corresponding initial value problem. As such, we have run numerical simulations of the full initial value problem (\cref{eq:TTW}) using Dedalus \citep{BurnsVOLB20} to verify our results. We use initial conditions corresponding to the initial iteration, $\tilde{\v{q}}_0$, and run each simulation until any transient oscillations have decayed. We find that converged steady states correspond to cases where strong frontal gradients do not form, and the system evolves over long timescales via shear dispersion. Conversely, cases where the non-linear iteration did not converge correspond to large Rossby number fronts in which strong secondary circulations drive the formation of strong horizontal gradients in the surface buoyancy profile which are eventually opposed by grid-scale horizontal mixing. Due to their similarity with results obtained using the iterative procedure, we do not show any results from numerical simulations here.

It should be noted that Dedalus simulations initialised in a strongly unbalanced state undergo strong inertial oscillations. These oscillations can form strong horizontal buoyancy gradients, even when the system would not form such gradients under more balanced initial conditions. This highlights an advantage of the iteration method; our results are not affected by strong, unphysical, transient evolution.

%Due to the need to include horizontal viscosity for stability when solving the IVP, it can be difficult to separate the effects of horizontal and vertical mixing. 

%This highlights an advantage of the iteration approach; there are no horizontal mixing processes--either from numerical diffusion or imposed viscosity/diffusivity--so the arrest of frontogenesis must be purely due to vertical processes. 

\section{Results}
\label{sec:results}

\begin{figure}
\centering
\includegraphics[width=0.6\linewidth]{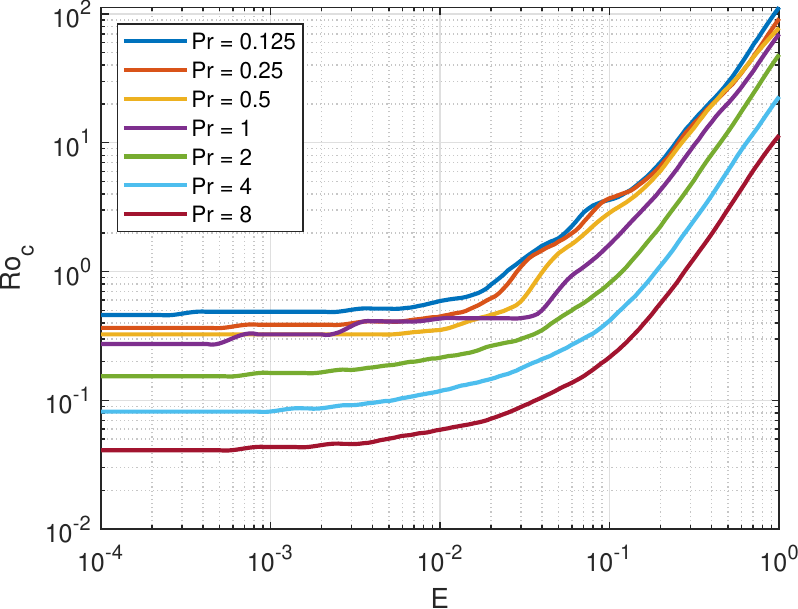}
\caption{Plot of the critical Rossby number, $\Ro_c$, as a function of Ekman number, $\E$, for different values of the Prandtl number, $\Pr$.}
\label{fig:critical_Ro}
\end{figure}

Our iteration procedure is run for Ekman numbers, $\E = 10^{-4} - 10^0$, and Rossby numbers, $\Ro = 10^{-2} - 10^2$. We find that steady state solutions exist for a given Ekman number, provided the Rossby number is below a critical value which we denote here as
\begin{equation*}
\Ro_c = \Ro_c(\E, \Pr). 
\end{equation*}
For  Rossby numbers above the critical Rossby number, $\Ro > \Ro_c$, the iteration procedure fails to converge and Dedalus simulations confirm the formation of sharp cross-front buoyancy gradients. \cref{fig:critical_Ro} shows the critical Rossby number as a function of Ekman number for several values of Prandtl number. As expected, we observe that stronger mixing (corresponding to larger Ekman numbers) enables steady state solutions to exist for larger values of Rossby number. Further, we observe that order $1$ values of $\Ro$ may have steady state solutions for realistic Ekman number values of $\E \sim 0.1$. For example, $Ro_c \approx 1.62$ for $(\E, \Pr) = (1,1)$.

From \cref{fig:critical_Ro}, we can determine the asymptotic dependence of $\Ro_c$ on $\E$ as
\begin{equation}
\Ro_c \sim \begin{cases}
1 & \E \ll 0.01,\\
\E^2 & \E \gg 0.01.
\end{cases}
\end{equation}
Further, we find that when the mixing of buoyancy is weak compared with the mixing of momentum ($\Pr > 1$), $\Ro_c$ decreases with increasing $\Pr$. This is consistent with observation from \citep{DauhajreEtAl2025} where they found that stronger gradients form in the limit of $\Pr \to \infty$, suggesting that the buoyancy mixing terms are purely frontolytic and hence frontogenesis must be driven by the secondary circulation. For $\Pr\lesssim 1$ the behaviour is complicated, with a non-monotonic dependence of $\Ro_c$ on $\Pr$ for some values of $\E$, likely arising from different balance regimes in the momentum equations. As turbulent Prandtl numbers are typically taken to be $1$, we do not explore this phenomena here. Finally, we note that $\Ro_c$ depends weakly on the form of $b_0$, however, changing $b_0$ to another appropriate profile (such as $b_0 = \tanh x$) does not qualitatively change our results or the parameter dependence discussed here.

\begin{figure}
\centering
\includegraphics[width=\linewidth]{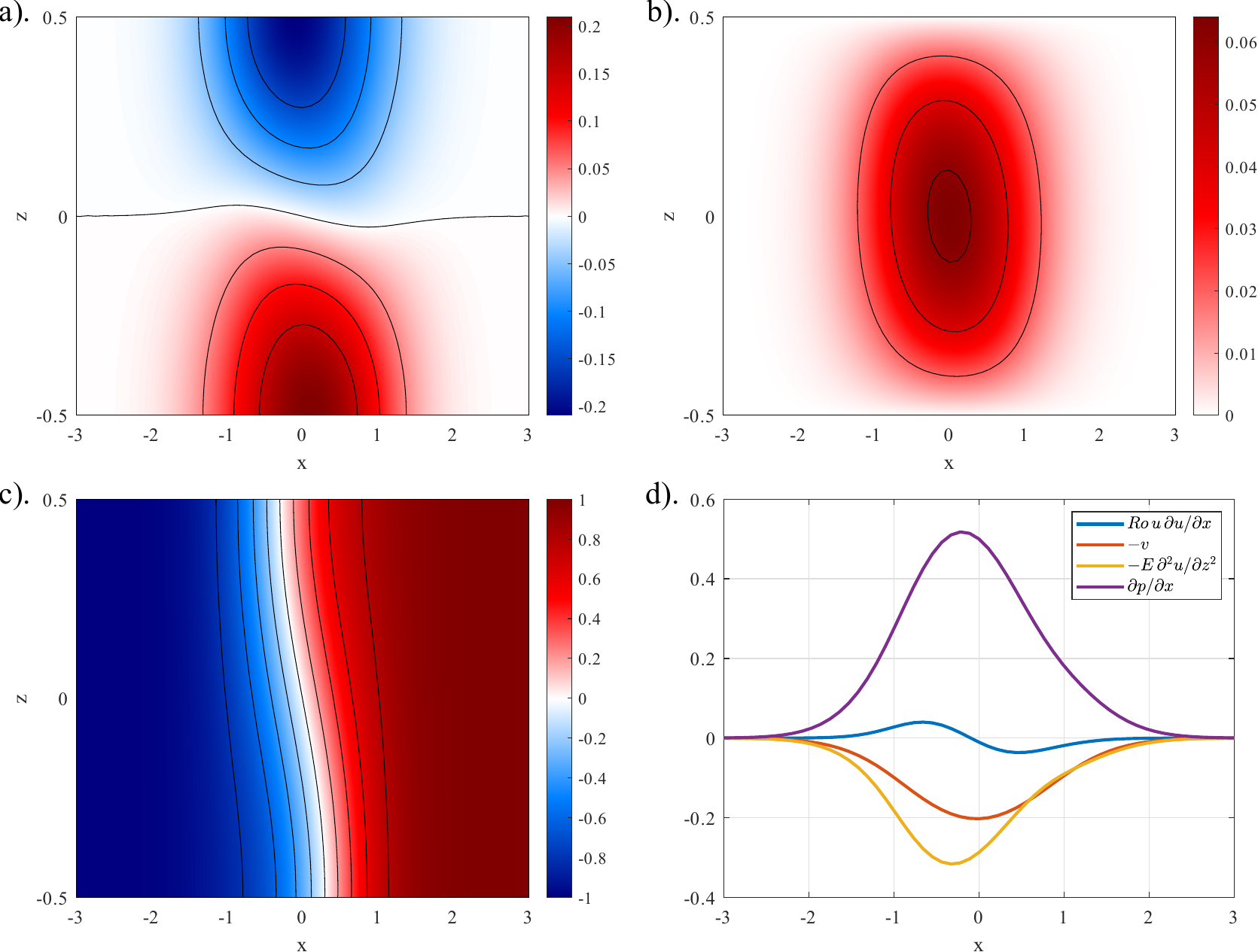}
\caption{Plots of the cross-front velocity, $u$, (a), circulation streamfunction, $\psi$, (b), and buoyancy, $b$, (c) as functions of $x$ and $z$. Panel (d) shows the terms from the cross-front momentum equation evaluated at $z = 0.5$ as functions of $x$. All results are shown for $Ro = 1.6$, $E = 0.1$ and $Pr = 1$.}
\label{fig:case_E_0.1_Ro_1.6}
\end{figure}

\cref{fig:case_E_0.1_Ro_1.6} shows the converged steady state solution for the case $(\Ro, \E, \Pr) = (1.6, 1, 1)$. Panel (a) shows the cross-front flow which acts to re-stratify the fronts and panel (b) shows the corresponding circulation streamfunction, $\psi$, defined as
\begin{equation}
(u, w) = \left(\pder{\psi}{z}, -\pder{\psi}{x}\right).
\end{equation}
Panel (c) shows the total buoyancy, consisting of the depth-averaged buoyancy component from \cref{eq:b0_def} and a stratified component arising from the tilting of the buoyancy contours by the secondary circulation. Panel (d) shows the terms from the cross-front momentum equation, evaluated at the top surface, $z = 1/2$. We observe that the system is in a state close to TTW balance, with the Coriolis force and vertical mixing term approximately balancing the cross-front pressure gradient. The non-linear advection term consists of a surface convergence on the low-buoyancy side of the front and is balanced by the vertical mixing term, which develops an asymmetric peak. This occurs as the secondary circulation is frontolytic, driving a horizontal flux of buoyancy towards the low buoyancy side of the front. The horizontal buoyancy flux is balanced by vertical mixing, which acts to mix away the surface convergence, thereby equilibrating the system. As the Rossby number is increased beyond $\Ro_c$, the vertical mixing is no longer sufficient to balance the frontogenetic tendency at the surface, sharp gradients form, and other physical processes must be invoked to prevent the formation of a surface singularity.

While the TTW solutions of \citet{CROWETAYLOR} are formally valid only for $\Ro \ll 1$, we can compare them with the steady state solutions obtained for cases where $\Ro = O(1)$. The total buoyancy buoyancy field prediction including terms up to $O(\Ro)$ is
\begin{equation}
b = b_0 - \Ro \Pr \sqrt{\E}\,K(\zeta) \left[\pder{b_0}{x}\right]^2 + O(\Ro^2),
\end{equation}
where $\zeta = z/\sqrt{\E}$ and $K$ is given in \cref{sec:app}. We may also extend the leading order asymptotic solutions for the horizontal velocity, $(u, v)$, to include $O(\Ro)$ corrections as
\begin{equation}
\left(u, v\right) = -\sqrt{\E}\,\big( K''\left(\zeta\right), K\left(\zeta\right)
\big) \pder{b_0}{x} +  \Ro\, \E\,\big(P_1(\xi),P_2(\xi)\big)\pder{}{x}\left(\pder{b_0}{x}\right)^2 + O(\Ro^2),
\end{equation}
where $P_1$ and $P_2$ are complicated functions of $\zeta$ given in \cref{sec:app}. Comparing these prediction with the near-critical steady state solution shown in \cref{fig:case_E_0.1_Ro_1.6} gives that the error in the prediction for cross-front buoyancy gradient, $M^2 = \pderline{b}{x}$, is around $4\%$, while the the error in the prediction for the velocity field is around $1\%$. This error reduces as $Ro$ is decreases away from the critical Rossby number $Ro_c$. This suggests that the asymptotic theory from \citet{CROWETAYLOR}---modified by the $O(Ro)$ velocity component in \cref{sec:app}---is sufficient to accurately describe these steady states, even outside it's formal range of validity.

\section{Discussion and Conclusions}
\label{sec:conc}

Here we have presented an idealised model of an ocean front that allows us to isolate the role played by vertical mixing in opposing the frontogenetic secondary circulation which arises in the case of finite Rossby numbers. We find that vertical mixing alone can prevent frontogenesis, provided the front is sufficiently weak; $\Ro < \Ro_c$ for some critical Rossby number $\Ro_c$. For stronger fronts, $\Ro > \Ro_c$, vertical mixing along is not sufficient to prevent the formation of a singularity in the surface buoyancy field. In this case, neglected processes such as horizontal mixing or shear instabilities are likely responsible for preventing unphysical frontogenesis.

Our critical Rossby number may be phrased in terms of a critical frontal width, $L_c$, as
\begin{equation}
L_c = \frac{1}{f}\sqrt{\frac{H\Delta b}{\Ro_c(E)}},
\end{equation}
and using typical frontal values of $H = 30\,\mathrm{m}$, $f=10^{-4}\,\mathrm{s}^{-1}$, $\Delta b = 10^{-3}\,\mathrm{ms}^{-2}$, $\nu = 0.01\, \mathrm{m}^2\mathrm{s}^{-1}$ gives a critical frontal width of $L_c \sim 3\,\mathrm{km}$. This value of $L_c$ is comparable to observed frontal widths, suggesting that vertical mixing may play an important role in arresting frontogenesis and setting the observed cross-front scale. In particular, vertical mixing likely acts to oppose strain-driven frontogenesis.

The method presented here uses non-linear iteration to find steady solutions and does not capture transient processes such as inertial oscillations, shear dispersion, and instabilities. As the system is susceptible to various instabilities for small Ekman number, our value for $Ro_c$ is likely an underestimate in the small $\E$ regime where mixing is enhanced by instabilities. In regimes with strong shear dispersion, this shear dispersion will act to widen the front, likely enhancing the frontolytic effects of vertical mixing and increasing the value of $\Ro_c$.

Our results have been verified using direct numerical simulations in Dedalus \citep{BurnsVOLB20}. We find that sub-critical cases, with $\Ro < \Ro_c$, closely match our iteration results. Conversely, super-critical cases, with $\Ro > \Ro_c$, rapidly form sharp surface gradients which are regularised by imposed horizontal dissipation. These simulations also highlight how inertial transients can lead to unphysical frontogenesis when initialising in an unbalanced state. As such, we echo the suggestion of \citet{AtkinsonEtAl2025} that care must be taken when choosing initial conditions for unstable or unbalanced system to avoid transient phenomena which may not be generalisable to real-world systems.

In addition to finding non-linear steady states, we have shown that the TTW solutions of \citet{CROWETAYLOR} may be extended to include $O(\Ro)$ velocity corrections which accurately describe the front---even for cases of order $1$ Rossby number---provided $\Ro < \Ro$. These analytical expressions for buoyancy and velocity can be used as balanced initial conditions in frontal simulations, or as a background state for studying frontal instabilities in the presence of vertical mixing. The study of instabilities in this framework is important for examining the interaction between enhanced turbulence and instability growth and will be an area of future work.

Throughout this study we have assumed that the imposed vertical mixing does not depend on time or the cross-front direction, limiting the applicability of our results to real-world fronts. While time-dependence cannot be described in this steady model, cross-front variation in the mixing can be easily included in our iteration method and is another avenue for future work.
\\\\
\noindent{\bf Code Availability\bf{.}} Matlab scripts implementing the non-linear iteration method and calculating the functions required for the analytical TTW solutions are available \href{https://github.com/mncrowe/TTW_Finite_Rossby_Number}{here}.
\\\\
\noindent{\bf Declaration of Interests\bf{.}} The author reports no conflict of interest.

\newpage
\appendix

\section{Higher Order TTW velocities}
\label{sec:app}

The TTW system in \cref{eq:BGS_mat} may be solved using an asymptotic expansion in Rossby number
\begin{equation}
f = f_0 + \Ro\,f_1 + \Ro^2\,f_2 +O(\Ro^3),
\end{equation}
for $f \in \{u,v,w,b,p\}$. The leading order solution is given in \citet{CROWETAYLOR} as
\begin{equation}
(u_0,v_0,w_0,p_0) = \left(-\sqrt{\E}\,K''(\xi) \pder{b_0}{x}, -\sqrt{E}\,K(\xi)\pder{b_0}{x}, \E\,K'(\xi)\pder{^2b_0}{x^2}, z b_0\right),
\end{equation}
where $b_0 = b_0(x)$ is the depth-averaged background buoyancy field and $\xi = z/\sqrt{\E}$. The function $K(\xi)$ is given by
\begin{equation}
K(\xi) = -\xi + A_+(\xi_0)\sin(\xi/\sqrt{2})\cosh(\xi/\sqrt{2}) + A_-(\xi_0)\cos(\xi/\sqrt{2})\sinh(\xi/\sqrt{2}),
\end{equation}
where $\xi_0 = 1/\sqrt{4\E}$ and
\begin{equation}
A_\pm(\xi_0) = \frac{\pm\sin(\xi_0/\sqrt{2})\sinh(\xi_0/\sqrt{2}) + \cos(\xi_0/\sqrt{2})\cosh(\xi_0/\sqrt{2})}{\sqrt{2}[\sin^2(\xi_0/\sqrt{2})\sinh^2(\xi_0/\sqrt{2}) + \cos^2(\xi_0/\sqrt{2})\cosh^2(\xi_0/\sqrt{2})]}.
\end{equation}
The $O(\Ro)$ corrections to the buoyancy and pressure fields may be easily calculated as
\begin{equation}
(b_1,p_1) = \left( -\Pr \sqrt{\E}\,K(\xi)\left(\pder{b_0}{x}\right)^2, \Pr\E\left(K'''(\xi)+\frac{\xi^2}{2}+c_1\right)\left(\pder{b_0}{x}\right)^2\right),
\end{equation}
where
\begin{equation}
c_1 = -\frac{1}{12\E} -2\sqrt{\E}\, K''(\xi_0).
\end{equation}
By substituting the $O(1)$ velocities and $O(\Ro)$ buoyancy and pressure into the $O(\Ro)$ horizontal momentum equations, the $O(\Ro)$ horizontal velocities are obtained as
\begin{equation}
(u_1,v_1) = \E\left(P_1(\xi),P_2(\xi)\right)\pder{}{x}\left(\pder{b_0}{x}\right)^2.
\end{equation}
Finally, the $O(\Ro)$ vertical velocity may be obtained through mass conservation as
\begin{equation}
w_1 = -\sqrt{\E^3} \left[ \int_{-\xi_0}^\xi P_1(\xi') \dint \xi'\right] \pder{^2}{x^2}\left(\pder{b_0}{x}\right)^2.
\end{equation}
%The functions $P_1$ and $P_2$ are complicated functions of $\xi$, $\xi_0$ and derivatives of $K(\xi)$ and are too long to include here. Full expressions are included as supplementary material and a Matlab script is available to evaluate these functions.
The structure functions $P_1$ and $P_2$ are given by
\begin{multline}
P_1(\xi) = A(K'(\xi)+1) + B K'''(\xi) + \frac{\Pr}{4}\left( \xi K''(\xi) + 2K'(\xi)+6 \right) + \frac{2}{5} K'^2(\xi)\\ -\frac{1}{2}K(\xi)K''(\xi) - \frac{1}{10} K'''^2(\xi) + \frac{1}{16} \xi K''(\xi)-\frac{1}{5} K'(\xi) - \frac{1}{16} \xi^2 K'''(\xi) - \frac{1}{10} - C_1, 
\end{multline}
and
\begin{multline}
P_2(\xi) = B(K'(\xi)+1) -A K'''(\xi) + \frac{\Pr}{4}\left( \xi K(\xi) + 3\xi^2 \right) - \frac{3}{10} K^2(\xi)\\ +\frac{1}{5}K''^2(\xi) - \frac{1}{2} K'(\xi) K'''(\xi) - \frac{1}{16} \xi^2 K'(\xi)-\frac{23}{80} \xi K(\xi) - \frac{1}{20} \xi^2 - C_2, 
\end{multline}
where the coefficients $A$ and $B$ are given by
%\begin{equation}
%A = \frac{K''(D)\left[(\frac{11}{80}-\frac{3\Pr}{4})K''(D)-\frac{D^2}{16} (K(D)+D)\right]+(K(D)+D) \left[(\frac{23}{80}-\frac{Pr}{4}) K(D)+\frac{D^2}{16} K''(D)+(\frac{1}{10}-\frac{3 \Pr}{2}) D\right]}{K''(D)^2+(D+K(D))^2},
%\end{equation}
%and
%\begin{equation}
%B = \frac{-(K(D)+D)\left[(\frac{11}{80}-\frac{3\Pr}{4})K''(D)-\frac{D^2}{16} (K(D)+D)\right]+ K''(D)\left[(\frac{23}{80}-\frac{Pr}{4}) K(D)+\frac{D^2}{16} K''(D)+(\frac{1}{10}-\frac{3 \Pr}{2}) D\right]}{K''(D)^2+(D+K(D))^2},
%\end{equation}
\begin{equation}
\begin{pmatrix}
A \\ B
\end{pmatrix} = \frac{1}{ K''^2+(\xi_0+K)^2} \begin{pmatrix}
K'' & K+\xi_0 \\ -(K+\xi_0) & K'' 
\end{pmatrix}
\begin{pmatrix}
(\frac{11}{80}-\frac{3\Pr}{4})K''-\frac{\xi_0^2}{16} (K+\xi_0)
\\
(\frac{23}{80}-\frac{Pr}{4}) K+\frac{\xi_0^2}{16} K''+(\frac{1}{10}-\frac{3 \Pr}{2}) \xi_0
\end{pmatrix},
\end{equation}
for $K$ and all it's derivatives evaluated at $\xi = \xi_0$. The constants $C_1$ and $C_2$ are given by
\begin{equation}
C_1 = \sqrt{\E} \int_{-\xi_0}^{\xi_0} K'^2(\xi) \dint  \xi,
\end{equation}
and
\begin{equation}
C_2 = \sqrt{\E} \int_{-\xi_0}^{\xi_0} K''^2(\xi) \dint  \xi +\Pr\left(2\sqrt{\E} K''(\xi_0)+\frac{1}{24\E}\right),
\end{equation}
and ensure that the functions $P_1$ and $P_2$ depth-average to zero.

The $O(\Ro^2)$ buoyancy may be obtained from the $O(\Ro^2)$ buoyancy equation as
\begin{equation}
b_2 = \E\Pr Q_1(\xi) \pder{}{x}\left(\pder{b_0}{x}\right)^3,
\end{equation}
where the structure function $Q_1$ for the $O(\Ro^2)$ buoyancy satisfies
\begin{equation}
Q_1''(\xi) = \Pr\left( \frac{2}{3} K''(\xi) K(\xi) - \frac{1}{3}K'^2(\xi) + C_1 \right) + \frac{2}{3} P_1(\xi),
\end{equation}
subject to conditions $Q_1(\pm \xi_0) = 0$. While an analytic form for $Q_1(\xi)$ can be found, it is generally easier to evaluate $Q_1(\xi)$ numerically by integrating this expression twice subject to the given boundary conditions.

Matlab scripts are available to evaluate these functions for a given vector of positions $z$ and parameters $(\E,\Pr)$.

\bibliographystyle{jfm}
\bibliography{bibliography}

\end{document}